\documentclass[pdflatex,sn-mathphys-num]{sn-jnl}

\usepackage{graphicx}%
\usepackage{multirow}%
\usepackage{amsmath,amssymb,amsfonts}%
\usepackage{amsthm}%
\usepackage{mathrsfs}%
\usepackage[title]{appendix}%
\usepackage{xcolor}%
\usepackage{textcomp}%
\usepackage{manyfoot}%
\usepackage{booktabs}%
\usepackage{algorithm}%
\usepackage{algorithmicx}%
\usepackage{algpseudocode}%
\usepackage{listings}%
\usepackage[most]{tcolorbox}
\usepackage{subcaption}
\usepackage{wasysym}

\theoremstyle{thmstyleone}%
\theoremstyle{thmstyletwo}%

\theoremstyle{thmstylethree}%

\begin{document}

\title[Generative Exaggeration in LLM Social Agents: Consistency, Bias, and Toxicity]{Generative Exaggeration in LLM Social Agents: Consistency, Bias, and Toxicity}


\author*[1]{\fnm{Jacopo} \sur{Nudo}}\email{jacopo.nudo@uniroma1.it}

\author[2]{\fnm{Mario Edoardo} \sur{Pandolfo}}\email{marioedoardo.pandolfo@uniroma1.it}
\equalcont{These authors contributed equally to this work.}

\author[2]{\fnm{Edoardo} \sur{Loru}}\email{edoardo.loru@uniroma1.it}
\equalcont{These authors contributed equally to this work.}

\author[1]{\fnm{Mattia} \sur{Samory}}\email{mattia.samory@uniroma1.it}

\author[1]{\fnm{Matteo} \sur{Cinelli}}\email{matteo.cinelli@uniroma1.it}

\author*[1]{\fnm{Walter} \sur{Quattrociocchi}}\email{walter.quattrociocchi@uniroma1.it}

\affil*[1]{\orgdiv{Department of Computer Science}, \orgname{Sapienza University of Rome}, \orgaddress{\street{Viale Regina Elena 295}, \city{Rome}, \postcode{00161}, \country{Italy}}}

\affil[2]{\orgdiv{Department of Computer, Control and Management Engineering}, \orgname{Sapienza University of Rome}, \orgaddress{\street{Via Ariosto 25}, \city{Rome}, \postcode{00185}, \country{Italy}}}


\abstract{We investigate how Large Language Models (LLMs) behave when simulating political discourse on social media. Leveraging 21 million interactions on X during the 2024 U.S. presidential election, we construct LLM agents based on 1,186 real users, prompting them to reply to politically salient tweets under controlled conditions. Agents are initialized either with minimal ideological cues (Zero Shot) or recent tweet history (Few Shot), allowing one-to-one comparisons with human replies.
We evaluate three model families—Gemini, Mistral, and DeepSeek—across linguistic style, ideological consistency, and toxicity. We find that richer contextualization improves internal consistency but also amplifies polarization, stylized signals, and harmful language. We observe an emergent distortion that we call ``generation exaggeration'': a systematic amplification of salient traits beyond empirical baselines.
Our analysis shows that LLMs do not emulate users, they reconstruct them. Their outputs, indeed, reflect internal optimization dynamics more than observed behavior, introducing structural biases that compromise their reliability as social proxies. This challenges their use in content moderation, deliberative simulations, and policy modeling.}

\maketitle

\section*{Introduction}
The generative capacity of Large Language Models (LLMs) is currently used in a variety of applications, including customer support, search engines, educational tools, and content moderation. In many cases, LLMs do not just operate in isolation but are embedded in systems that require interaction or behavioral consistency.
This shift—commonly referred to as agentification—consists of employing LLMs as agents that autonomously execute actions in interactive settings \cite{debenedetti2024agentdojo, boskabadi2025industrial,shen2023hugginggpt,goodell2025large,m2024augmenting,kim2025artificial}. Agentified LLMs may perform social roles, sustain dialogue, and influence information flows within dynamic environments \cite{coppolillo2024engagement, park2023generative}.

Unlike traditional agent-based models \cite{epstein1996growing,conte2001sociology}, which explicitly codify rules and assumptions to describe how individuals make decisions and interact, LLM-based agents generate behavior based on patterns extracted from large-scale training data. As a result, they may inherit the statistical biases present in the data and be influenced by its training procedure, rather than by any theoretical structure. This could have important consequences for how such agents behave, and how their outputs should be interpreted and evaluated.

In the near future, the prevalence of agentified LLMs is expected to expand \cite{moller2025impact}, with LLMs taking on delegated roles in contexts where behavioral plausibility and social coherence matter. To understand the implications of this trend, it is necessary to examine what types of behavior LLMs produce and which distortions they may introduce into the environments in which they operate.

Social media platforms are a crucial testing ground to assess this shift. Their architecture already favors amplification of emotionally charged and polarizing content, often distorting political discourse and reinforcing tribal identities \cite{cinelli2021echo, di2024characterizing, del2016spreading, donkers2025understanding}. Using LLMs to simulate users in settings where political discussion happens in real time raises practical questions. What happens when generative agents are tasked with simulating human users engaged in political debate? Do they faithfully preserve the behavioral patterns of the profiles they emulate, or introduce new, systematic distortions?

In this study, we investigate the use of LLMs to simulate real users involved in the 2024 U.S. presidential election on X (formerly Twitter). 
We implement over 1,000 synthetic agents using six language models from three different model families---Gemini, Mistral, and DeepSeek---evaluating both larger and smaller variants to examine size-related effects. The three main model families that we study come from different geographical regions---the US, China, and Europe, respectively---which may result in differences in how they handle politically sensitive topics or interpret specific cultural references.  Such a comparative approach aims to examine whether all models exhibit consistent patterns or if variations may emerge as a consequence of differences in technical capabilities, training data, and geopolitical development environment.

Each agent is initialized with varying degrees of behavioral context regarding the reference human user it simulates (e.g., political leaning, comment history, profile data) and prompted to reply to politically salient posts. We analyze LLMs' output along three axes: linguistic consistency, political leaning, and toxicity.

Our findings reveal clear patterns of a phenomenon that we refer to as \textit{generative exaggeration}. 
Additional context improves linguistic coherence and ideological consistency with the reference users, but it simultaneously increases ideological extremism and verbal toxicity. Specifically, agents initialized with minimal context often fail to emulate real users faithfully. When provided with richer behavioral traces, the simulation improves in coherence but deteriorates in realism, producing stereotypical and exaggerated portrayals. This includes the overuse of partisan hashtags, emojis, and emotionally charged phrasing, as well as increased toxicity and partisan animosity. Importantly, this distortion is not symmetric: we find a consistent tendency to caricature right-leaning users more than left-leaning ones, though both are affected.

These results suggest that simulating political behavior with LLMs may not be a neutral process: LLM operators are tasked with solving the paradox that feeding more information and increasing computational performance may exaggerate certain traits and produce less realistic outputs, raising concerns about the use of LLMs in high-stakes contexts such as political communication and democratic deliberation. 

This result adds to a growing literature using LLMs to simulate political personas \cite{koley2025salm, rossetti2024social, cau2024bots}, run polling experiments \cite{holtdirk2024fine}, and test moderation strategies in controlled settings \cite{park2023generative, park2022social}. This body of work showed how models can exhibit emergent patterns over time, such as reinforcing bias or converging toward specific opinions \cite{ahnert2025simulating, coppolillo2025unmasking, taubenfeld2024systematic}, highlighting the importance of developing strategies to best align LLMs with users' behavior.
However, the side effects of increasing simulation fidelity, such as the trade-off between staying close to a user’s behavior and exaggerating some of its features, have not been studied in detail. While prior work has explored the biases of LLM-generated personas \cite{cheng2023compost, Liu2024, Li2025-op,alipour2024robustness}, we know little about how these models engage in political communication when simulating replies based on real user data.

Several studies have raised concerns about the tendency of LLMs to reflect or amplify societal biases \cite{hu2025generative, loru2025decoding}, including geopolitical bias \cite{buyl2024large, noels2025large} and alignment with specific ideological narratives \cite{chen2024susceptible}. Other work has looked at the generation of hate speech in synthetic content \cite{civelli2025impact}, or at how model outputs can reinforce polarization in subtle ways \cite{piao2025emergence, sharma2024generative}. More generally, how LLMs shape digital conversations and structure discourse remains an open research question \cite{lazovich2023filter}. The present work shows how optimizing for fidelity along one user trait---political leaning---may spill over to unintended domains like verbal toxicity. 

In this context, we make three main contributions. First, we provide empirical evidence on how LLMs simulate political users, showing systematic differences between the generated and original behavior. Second, we introduce the concept of generative exaggeration, a structural tendency of LLMs to amplify salient traits when simulating individuals. Third, we show that exaggeration affects political profiles asymmetrically.
Our analysis suggests that these effects are not driven by prompt design but are a byproduct of model training and optimization. As a result, using LLMs as social agents—especially in sensitive contexts such as political communication—requires careful evaluation of the structural biases these systems may introduce.

\section*{Results and Discussion}

To assess differences between human users and LLM, we analyze the behavior of LLM-based agents simulating politically active users on social media. The objective is to quantify the extent to which large language models replicate—and distort—human behavior in online political discourse.

To this end, we use a public dataset of over 21 million interactions on X (formerly Twitter) related to the 2024 U.S. presidential election \cite{balasubramanian2024public}. This corpus enables the reconstruction of user profiles in a real-world context where political identity is salient and interactions are high-stakes.

We focus on tweet–reply interactions authored by 1,186 users with at least 50 prior tweets. This threshold ensures a sufficient behavioral signal to estimate political leaning and reliably capture individual linguistic style, while maintaining a large enough sample size for comparison. Each agent is thus prompted to respond to the same tweet as the original user, enabling direct tweet-level comparisons across multiple behavioral dimensions, including lexical diversity, ideological consistency, and toxicity of language. 

To enable comparisons across the ideological spectrum, each user is assigned a political leaning score. We compute this score by applying a stance classifier to at least 50 tweets per user, labeling each message as pro-Democrat, pro-Republican, or neutral. Then, we assign a numerical value to each label (+1 for Republican, –1 for Democrat, 0 for Neutral) and average these values across the user’s tweets. Based on the resulting score, users are categorized as Democrat, Neutral, or Republican. User- and LLM-generated responses are classified through the same procedure. Further methodological details are provided in the Methods section.

Agents are tested under two initialization conditions. In the Zero Shot setting, the model receives only the user’s inferred political leaning. In the Few Shot setting, the model is provided with the user’s nickname, bio, and a sample of their past tweets. The specific prompting strategies used in each case are described in the Methods section.

Our analysis is guided by four research questions, which we tackle in the following sections:

\begin{enumerate}[]
\item \textbf{Lexical Realism.} \textit{Do LLM agents differ from humans—and from each other—in expressive style when simulating political users?}
\item \textbf{Ideological Consistency.} \textit{How accurately do LLMs reproduce users’ political leanings, and how does fidelity vary by prompt conditioning?}
\item \textbf{Toxicity Amplification.} \textit{Do LLMs generate more toxic content than humans, and how does this depend on prompting strategy and political identity?}
\item \textbf{Generative Exaggeration.} \textit{Do LLMs systematically amplify salient user traits, and under what conditions does this distortion emerge?}
\end{enumerate}

\subsection*{Lexical Realism and Stylization}

We start our analysis by assessing the ability of LLM-based agents to emulate human behavior at a linguistic level. First, we examine potential differences in length between tweets written by humans and by agents. This is an important consideration, as lexical measures are inherently sensitive to the length of the input text. 
Our results show that, although no model perfectly reproduces the distribution observed in human-authored tweets, the lengths of generated tweets generally fall within a plausible range, typically on the order of $10^2$ characters, with the majority remaining under Twitter’s 280-character constraint. 
However, we also identify a small fraction of anomalous tweets that exceed the maximum character length. These tweets are most often generated by the smaller models and are statistical outliers.
Consequently, these instances were excluded from all subsequent lexical analyses. The tweet length distributions are displayed in full in Supplementary Figure S4, while a detailed breakdown of all removed tweets is reported in Supplementary Table S1.

We evaluate lexical diversity by computing the Type-Token ratio (TTR), which quantifies the proportion of unique words (types) to the total number of words (tokens) in the text. This measure allows for a straightforward interpretation: a higher TTR reflects a more varied vocabulary, while a lower TTR indicates frequent repetition of the same words. TTR-based measures are well-established in the literature as indicators of lexical variation and textual complexity~\cite{di2024patterns, tweedie1998variable, mccarthy2010mtld, rosillorodes2025entropytypetokenratiogigaword, richards1987type, kettunen2014can}. In particular, we focus on the LogTTR, or Herdan’s C~\cite{herdan1960type, chotlos1944iv, Weitzman_1971}, a variant of the standard TTR that is more robust to varying text lengths. The detailed formulation of this measure, as well as all text preprocessing steps performed, are reported in Methods.

\begin{figure}[t] 
    \centering
    \includegraphics[width=\textwidth]{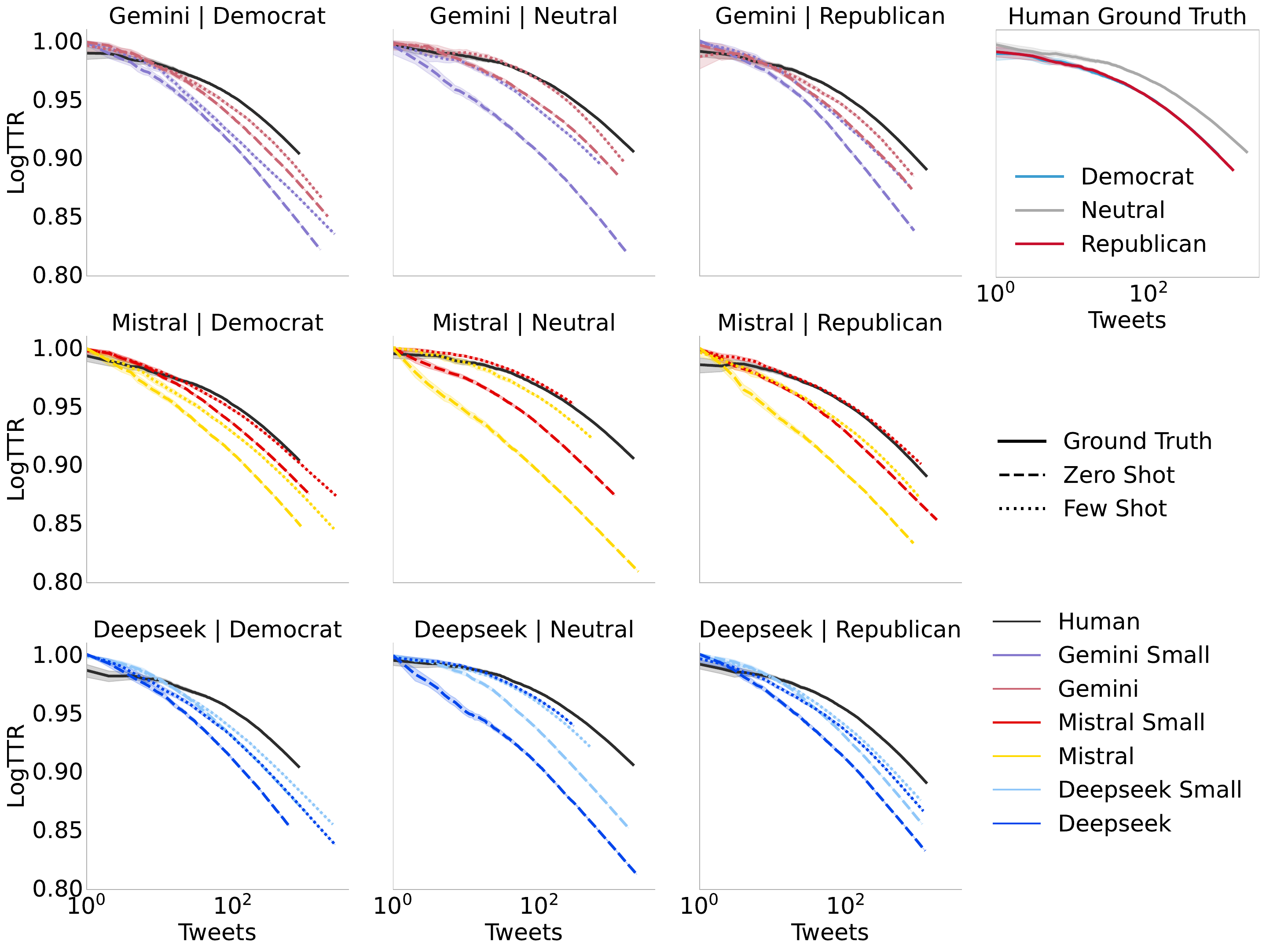} 
\caption{LogTTR in the case of human and model-generated tweets, across different LLMs (rows) and political leanings (columns). Each panel shows how lexical diversity (LogTTR) evolves as more tweets are added (log-scaled). The curves represent the average over 100 simulations, each based on a different random ordering of tweets, to smooth variability and highlight general trends. Shaded areas indicate 95\% confidence intervals, computed from 1000 bootstrap resamples. The Few Shot strategy appears to help models better approximate real human lexical behavior, bringing their output closer to human-authored content in terms of lexical diversity. In contrast, the top-right panel presents the LogTTR behavior of human-generated tweets exclusively, conditioned by political leaning. This analysis reveals that Democrat and Republican users exhibit comparable lexical diversity trends, while Neutral users demonstrate higher diversity.}
    \label{fig:incremental_logttr_score}
\end{figure}

In Fig.~\ref{fig:incremental_logttr_score}, we show how lexical diversity, measured using LogTTR, evolves as tweets are sequentially added to a growing corpus. The analysis is presented separately for each model and political leaning to assess any potential difference. By construction, all resulting curves are monotonically decreasing, since adding more text generally increases the number of unique words at a slower rate than the total number. Hence, in this analysis, we are interested in comparing how these curves decrease relative to one another across models and political leanings, as well as against human-generated tweets.

The panel in the top-right corner shows the LogTTR curves of human tweets alone, grouped by political leaning. Democrat and Republican users show similar levels of lexical diversity. In contrast, Neutral users tend to have higher diversity. This likely reflects the fact that non-partisan content covers a wider range of topics and vocabulary, whereas partisan tweets may often repeat similar terms, slogans, or rhetorical patterns.

Comparing human tweets with LLM-generated replies, we observe that all models tend to produce higher LogTTR scores than humans when considering small tweet samples, particularly for tweets inferred as Republican- or Democrat-leaning. However, this trend reverses when the number of examined tweets increases, as model-generated content begins to exhibit lower LogTTR values compared to human-authored tweets. This pattern indicates that although LLMs appear lexically rich over a small number of tweets, their vocabulary becomes increasingly repetitive as a larger corpus is taken into account.

The curves exhibit different end points due to the fact that a model may generate a tweet that reflects a political leaning different from that of its ground truth counterpart. In several cases, models show greater deviation from the ground truth when simulating Neutral users, though the extent varies across model families and prompting strategies. This political leaning category is underrepresented in the generated data due to the fact that most political shifts occur within it.

Initialization strategy and model size both affect the lexical diversity manifested by agents. The Few Shot strategy, in particular, shifts the LogTTR curves of agent-generated content closer to those of real users. When comparing model sizes, the large variant of Gemini better approximates human lexical diversity than its smaller counterpart. In contrast, for the Mistral and DeepSeek families, the smaller models produce outputs more consistent with human behavior. Among all configurations, Mistral Small with Few Shot initialization produces the LogTTR curves that most closely match those of human users across all political leanings.

Repeating the analysis using the standard TTR metric yields results that are consistent with those obtained from LogTTR (see Supplementary Figure S4). A summary of the corresponding quantitative analysis of the texts is provided in Supplementary Table S2.

These findings suggest that, while current LLMs can reproduce surface-level lexical variation, their outputs are still influenced by internal priors that diverge from those observed in human behavior. This indicates inherent constraints in the capacity of LLMs to faithfully simulate real users, even under optimized initialization strategies.

\subsection*{Ideological Bias and Consistency}

\begin{figure}[t] 
    \centering
    \includegraphics[width=\textwidth]{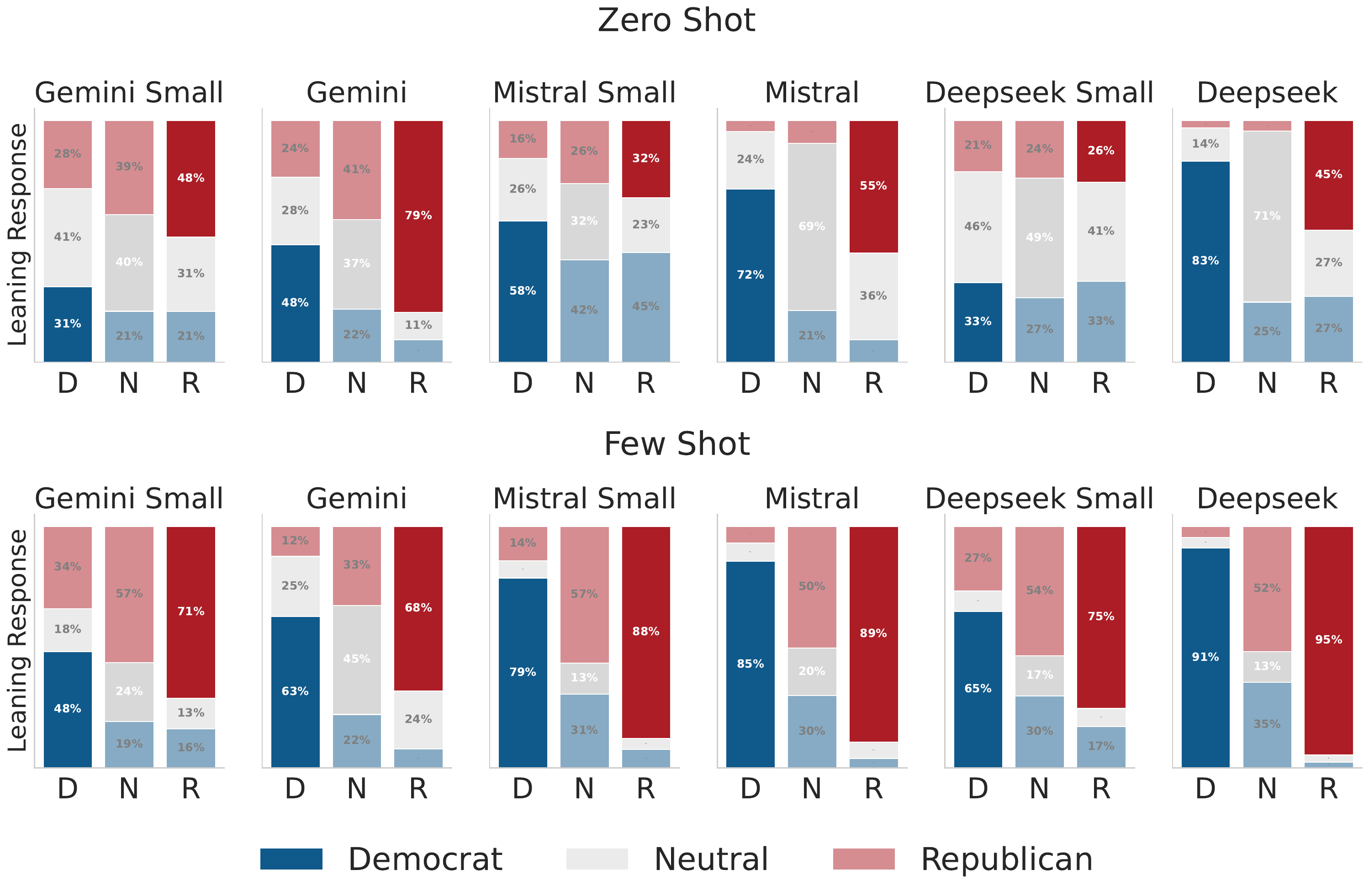}
\caption{ For each user simulated via an LLM-based agent, we evaluate the political leaning of the generated tweets (y-axis) with respect to the user's original leaning class (x-axis), and report the percentage of tweets that are consistent with the initialized political leaning. When models are initialized in a Zero Shot manner—using only the user's political leaning—they tend to produce politically neutral outputs, lacking fidelity to the intended ideological stance. However, when Few Shot prompting is used to provide contextual examples, models more accurately embody the target political identity, generating comments that better reflect the intended leaning. This increased alignment comes at the cost of decreased neutrality for users labeled as politically neutral.}
    \label{fig:leaning_cond_prob}
\end{figure}

In this section, we assess whether agent responses align ideologically with the users they are meant to simulate. Following previous work \cite{rozado2024political, rottger2024political}, we analyze the inferred political leaning of the generated replies. Agents are first grouped according to the political leaning of their target user—Democrat, Neutral, or Republican. We then apply an automated classifier \cite{burnham2024political} to label each generated reply and compute the conditional probability of producing a reply with a given leaning, given the emulated user's profile. Note that in the Zero Shot condition, agents are explicitly given the user’s political leaning, whereas in the Few Shot condition, this information is not directly provided unless included in the user's tweets or bio.

Figure~\ref{fig:leaning_cond_prob} shows the distribution of political leanings in tweets generated by agents, grouped by the political leaning of the corresponding users. We observe a clear difference across prompting strategies. 
In the Zero Shot setting, where agents are only prompted with the user's political leaning, no clear pattern emerges. Most agents tend to produce replies across the whole political spectrum, regardless of the political leaning they were provided at initialization. In contrast, Few Shot prompting, which is based on a user’s tweet history, results in responses more consistent with the user’s original stance, but at the cost of decreased neutrality. Notably, users originally labeled as Neutral are more often emulated as more politically aligned than they actually are, suggesting a context-driven distortion in how LLMs capture political identity. 

For instance, when prompted to simulate a Republican-leaning user, DeepSeek generates ideologically aligned responses in approximately 45\% of cases in the Zero Shot condition. However, this proportion rises to 95\% in the Few Shot setting, indicating an improved consistency with the reference ideology. At the same time, when focusing on users labeled as Neutral, we observe that the proportion of generated tweets drops from 71\% in the Zero Shot setting to 13\% in the Few Shot setting. This shift highlights a tendency of LLMs to drift toward polarized outputs even when exposed to minimal ideological cues. An exception is the larger Gemini model, which exhibits a high proportion of partisan comments even in the Zero Shot setting, as well as a sizable presence of neutral comments in the Few Shot setting.

\begin{figure}[t]
    \centering
\includegraphics[width=\textwidth]{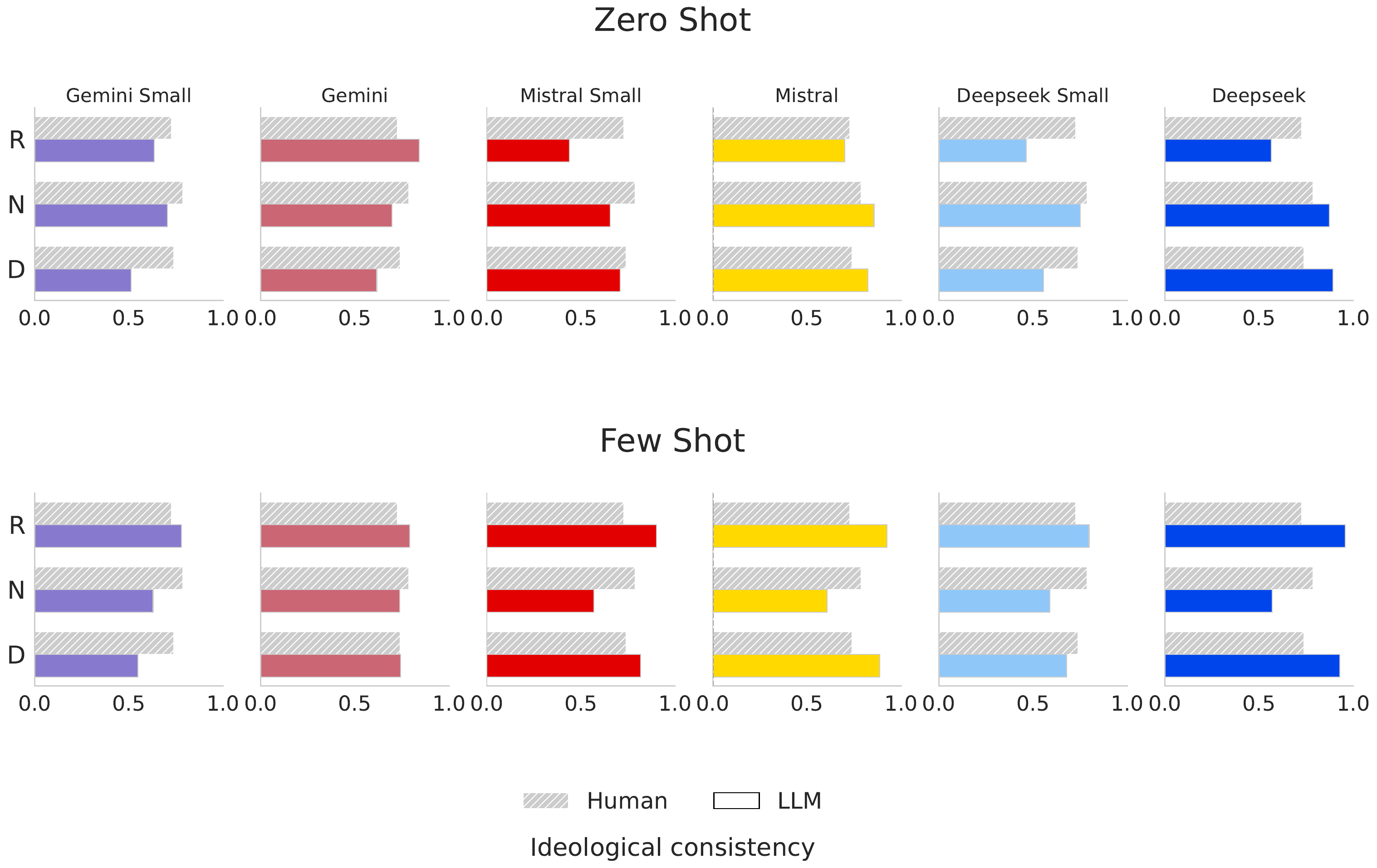} 
    \caption{
Ideological consistency for each model for the two initialization strategies (Zero Shot or Few Shot). The metric ranges from $0$ to $+1$ and measures the coherence between the agent's initial political leaning and that of the generated comments. See Eq. \eqref{eq:id_fid_gap} in Methods for its formal definition.
A value close to $1$ indicates a caricatured ideological portrayal by LLMs and that all responses contain political content expressing the subject’s political leaning. Conversely, a value close to $0$ results from agents employing a less politically overt tone than their human counterparts. Overall, agents generate outputs that more accurately reflect the user's stance in the Few Shot setting, though there are exceptions, such as DeepSeek or Mistral.
}
    \label{fig:ideol_fid_gap}
\end{figure}

Figure~\ref{fig:ideol_fid_gap} illustrates the models' `ideological consistency', a metric we introduce to quantify the coherence between an individual’s expressed opinions (through comments) and their known ideological leaning. Specifically, it measures how closely the political leaning of generated replies aligns with the user's political leaning. To compute this, we first measure, separately for humans and agents, how closely each message aligns with the user's estimated political identity. We then average this alignment across all messages and users to obtain the `ideological consistency loss' $\mathcal L$. Finally, we define $\mathcal C^A = 1 - \mathcal L^A$ as the ideological consistency of agents and $\mathcal C^H = 1 - \mathcal L^H$ of humans. A formal definition of these metrics is provided in Methods.

The values of ideological consistency range from 0 to +1. A higher value indicates that the comments are more likely to reflect the true political leaning of the user or agent, demonstrating strong alignment between expressed opinions and ideological stance. Conversely, a value closer to zero suggests little or no alignment.

In the Zero-Shot setting, the ideological consistency of most agents simulating Republican users is lower than that observed in humans, suggesting that their responses are less ideologically aligned compared to their human counterparts. The larger variants of Gemini and Mistral are exceptions to this tendency. For Democrat and Neutral users, instead, we observe a larger variation across models. Notably, agents modeling Neutral users are generally characterized by an ideological consistency similar to that observed in humans. Mistral stands out as the model that most faithfully replicates the political stance of users across the spectrum.

When evaluating agents initialized with the Few Shot procedure, we find notable differences from the Zero Shot setting across all models, except for the two Gemini variants.
Overall, adding a user's comment history to the prompt tends to increase the ideological consistency of agent responses with the original users, as shown in Fig.~\ref{fig:leaning_cond_prob}. However, this increase in alignment is accompanied by other changes that may reduce the behavioral consistency of the outputs.
The generally high ideological consistency for both Democrat- and Republican-leaning users suggests that agents produce tweets that align more closely with the model’s own political leaning than the original users do. In this way, the models generate a caricatured representation of the user’s political leaning within the debate.
This is particularly evident for the larger Mistral and DeepSeek models, both displaying a substantially different behavior when compared to the Zero Shot case. In contrast, the larger Gemini model produces the agents that best reproduce the original users' political output in this setting. 

\subsection*{Toxicity}

To quantify the prevalence of toxic or harmful content in generated tweets, we annotate each reply using Perspective API \cite{lees2022new}, a widely adopted classifier for detecting toxic language in online content \cite{avalle2024persistent}. The API defines toxicity as ``a rude, disrespectful, or unreasonable comment likely to make someone leave a discussion'', and assigns a toxicity score ranging from 0 (non-toxic) to 1 (highly toxic). Specifically, we compute the proportion of tweets exceeding a toxicity score of 0.6, a threshold commonly adopted in prior work on online toxicity to label a message as `toxic' \cite{avalle2024persistent}.

\begin{figure}[t] 
    \centering
    \includegraphics[width=\textwidth]{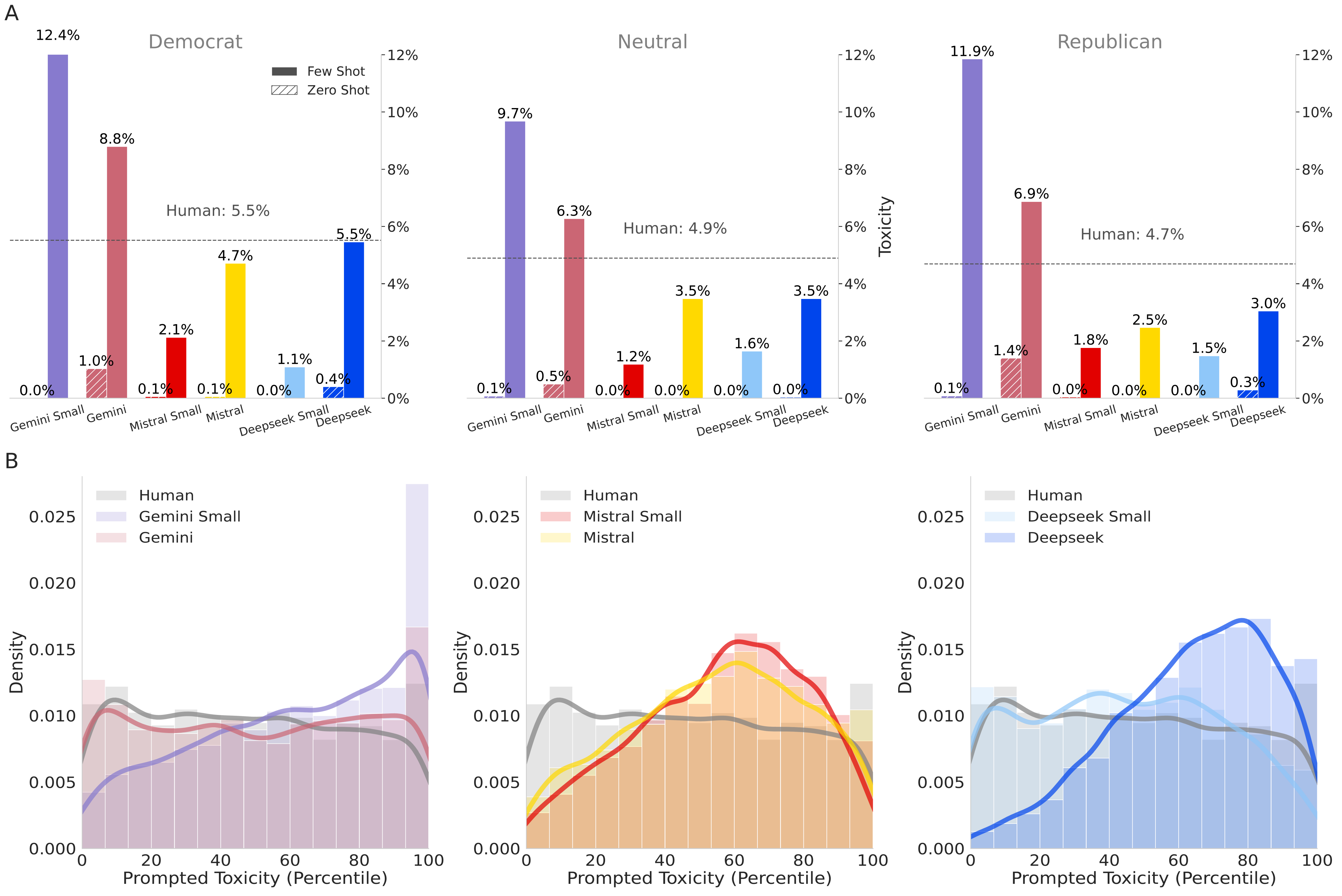} 
    \caption{
(A) Proportion of responses with toxicity score \(> 0.6 \), grouped by LLM, initialization method (Zero Shot and Few Shot), and political leaning.
Providing contextual comments results in higher toxic output, especially with models like Gemma and Gemini, both developed by Google, which exceed the human baseline average of 5\%.
(B)
Distribution of toxicity percentiles for synthetic tweets generated by different models. Each tweet’s toxicity percentile is calculated relative to the toxicity levels of the 30 comments in the prompt, indicating where the tweet's toxicity ranks within that reference set. While Mistral-based models tend to produce a roughly Gaussian distribution, Google's models exhibit a relatively uniform shape with a sharp peak at the 100th percentile, suggesting a tendency to exceed the toxicity level present in the prompt.
}
    \label{fig:toxicity}
\end{figure}

Leveraging these annotations, we assess both the prevalence and severity of toxic language in human replies and in their LLM-generated counterparts, evaluating the degree to which such behavioral traits are faithfully preserved or distorted in the simulation. We apply this metric across all agent replies, comparing model outputs to human baselines. Additionally, we group the results by the political leaning class used to prompt the agents, allowing us to examine whether toxicity levels vary systematically across the political spectrum.

As shown in Fig.~\ref{fig:toxicity}A, the prompting strategy (Zero or Few Shot) appears to play a crucial role in the toxicity levels exhibited by the agents. When agents are initialized using the Zero Shot procedure, they produce a small fraction of toxic replies across all LLMs under evaluation. Users, in turn, produce approximately $5\%$ toxic comments. 
When agents are provided with a set of user comments, in the Few Shot setting, toxicity does emerge. This is particularly evident for Gemini models, which, in this setting, frequently generate toxic content, even exceeding human reference levels. 

Toxicity can appear in LLM agents when it exists in the tweets used as prompts, even if it is not the typical behavior of the models. Models generally respect safety constraints in the Zero Shot setting, but Few Shot prompting---which adds more user context---can weaken these controls and allow harmful content. This trade-off poses a challenge for using LLMs in social simulations: improving behavioral accuracy may reduce safety measures.

Building on these findings, to assess whether LLM agents reflect or amplify the toxicity levels present in their prompting context, we compare the toxicity of generated responses with the percentile distribution of toxicity in the user’s historical tweets (i.e., the Few Shot prompt strategy). This approach allows us to evaluate whether agents reproduce the observed variability in human behavior or exhibit deviations. Deviations from the user's baselines, such as consistent overproduction of toxic content relative to the provided tweets, would signal the emergence of systematic exaggeration. 

Formally, consider a tweet with toxicity score $t$ and a set of $M$ reference tweets with toxicity scores $\{r_j\}_{j=1}^M$. The percentile rank $P(t)$ is defined as the proportion of reference tweets with toxicity less than or equal to $t$:

\begin{equation}
    P(t) = \frac{1}{M} \sum_{j=1}^M \mathbf{1}\big(r_j \leq t \big)
\end{equation}

where $\mathbf{1}(\cdot)$ denotes the indicator function. In our specific setup, we set $M=30$, which corresponds to the number of tweets provided to each agent for the Few Shot initialization. 

As shown in Fig.~\ref{fig:toxicity}B, nearly all models tend to \textit{overshoot} the reference toxicity range of the users they simulate: the distribution of toxicity scores in the generated replies has a center of mass above the 50th percentile of the prompt distribution in most cases. This effect is particularly pronounced in the larger model variants, which often generate outputs more toxic than the tweet samples they were initialized with.
For instance, larger versions of Mistral and DeepSeek produce left-skewed distributions, indicating a consistent upward shift in toxicity. Gemini, in contrast, displays a sharply peaked distribution—especially in its smaller variant—suggesting that the generated replies cluster around the most toxic subset of the prompt tweets. This pattern reveals a generative bias toward amplifying the more extreme or emotionally charged elements of the input, rather than sampling proportionally from the full distribution of prior user behavior.

\subsection*{The Style of Generative Exaggeration}

Our findings suggest that LLMs do not faithfully reproduce the users they are meant to simulate, across several key behavioral attributes such as ideology and toxic language. Instead, they tend to amplify specific prominent traits, especially in the Few Shot setting, resulting in a systematic distortion of user behavior. This reflects a form of \textit{generative exaggeration}, in which the agent captures superficial markers of identity while misrepresenting the underlying behavioral profile and creating a caricatured portrayal. This behavior is not incidental. It reflects a tendency of models to prioritize linguistic salience and consistency over contextual depth. When models are conditioned on partisan cues, they do not infer ideology---they generate responses based on prominent patterns in the input data. Hence, fidelity collapses into caricature.

We now aim to better characterize this phenomenon, shifting our attention toward the use of emojis and hashtags in the agent-generated replies. This is particularly relevant in the context of political discourse, as these elements are typically employed as markers of ideological partisanship, especially in a polarized setting.

\begin{figure}[t]
    \centering
    \includegraphics[width=1\textwidth]{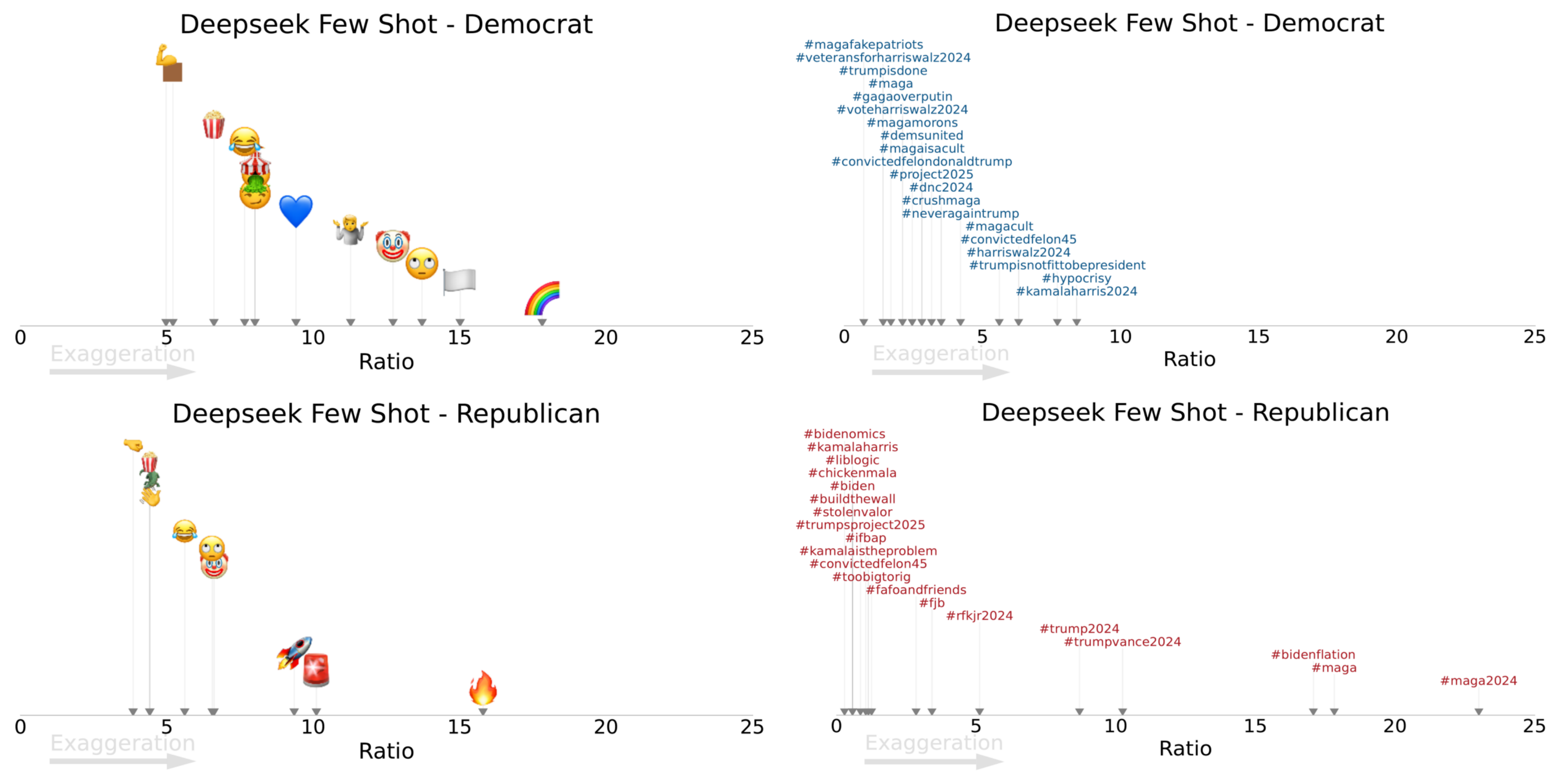} 
    \caption{We analyze the use of emojis and hashtags in tweets classified as Democratic or Republican, comparing human and LLM-generated content. The arrow labeled ``Exaggeration'' starts on the x-axis at a ratio of 1 and marks the beginning of LLM overrepresentation. In Panel (A), each point represents an emoji used either by humans or by models. The x-axis shows the ratio between the relative frequencies—defined as the fraction of tweets containing the emoji—of LLM-generated versus human-generated content, computed separately for Democratic and Republican tweets. This visualization highlights emojis that are disproportionately used by the model compared to humans. Models tend to overuse expressive and stereotypical emojis, such as the clown face, laughing face, and confused face. Notably, the rainbow (associated with Democratic discourse) is more frequent in model-generated Democratic tweets, while the fire or the rocket appear more in Republican ones. Panel (B) replicates the same analysis for hashtags: the x-axis again represents the ratio between model and human relative frequencies. The results reveal a consistent overuse of politically charged hashtags by the model across both political alignments, with a slight overrepresentation in Republican tweets.}
    \label{fig:deepseek_emot_hash}
\end{figure}

Further characterizing the results presented in Fig.~\ref{fig:ideol_fid_gap}, human-authored tweets contain emojis and hashtags relatively rarely—87.47\% and 96.29\% of tweets, respectively, omit them. In contrast, agents include these elements much more frequently, especially in the Few Shot setting. For instance, Mistral-generated replies contain emojis and hashtags in over 49\% of cases. See Supplementary Table S2 for a full breakdown. 

To evaluate the extent of this amplification, Fig.~\ref{fig:deepseek_emot_hash} shows the proportion of human- and agent-generated tweets containing each emoji or hashtag present in both sets. In the figure, a ratio of 1 indicates an equal number of occurrences in both sets, while a ratio greater (smaller) than 1 is a marker of LLM overrepresentation (underrepresentation). We observe that overall, emoji use is markedly exaggerated, especially for those linked to stereotypical political identities. For example, the rainbow emoji---typically associated with progressive viewpoints---appears nearly 20 times more often in DeepSeek-generated tweets than in authentic human replies. A comparable inflation is observed in hashtag usage. While some ideologically charged hashtags (e.g., \textit{\#kamalaistheproblem} or \textit{\#oldmantrump}) are slightly underrepresented, others are substantially amplified. For instance, hashtags like \textit{\#kamalaharris2024}, \textit{\#MAGA}, and \textit{\#MAGA2024} appear up to 10-15 times more frequently in LLM outputs than in real user content. Although Fig.~\ref{fig:deepseek_emot_hash} focuses on DeepSeek in the Few Shot setting, comparable behaviors are found across all evaluated models (see Supplementary Figures S5 and S6 for hashtags and emojis, respectively).

The over-production we observe is systematic, not random. During training, LLMs minimise next-token error by giving extra weight to high-salience ideological tokens---slogans, labels, polarising phrases---because those tokens strongly predict surrounding text. The generated output therefore looks ideologically coherent, yet it collapses rich political positions into a few over-used symbols. If such models take over content moderation or deliberation, this skew becomes part of the decision pipeline: instead of merely routing discourse, the agents reshape it by amplifying loud cues and muting low-frequency nuance.

\section*{Conclusions}

This study investigated how large language models behave when tasked with simulating users engaged in online political discourse. We constructed over 1,000 synthetic agents modeled on real users involved in the debate on X surrounding the 2024 U.S. presidential election, prompting them to reply to politically salient content under controlled experimental conditions. Agents were initialized either with a Zero Shot procedure, receiving only the user's estimated political leaning, or a Few Shot procedure, where they are provided with the user’s bio and message history.

Across four key dimensions---lexical diversity, ideological consistency, toxicity, and stylistic exaggeration---our results show a systematic pattern: LLMs reshape user behavior in ways that reflect structural biases, rather than simply mirroring it. In the Zero Shot setting, responses are distributed uniformly across the political spectrum, regardless of the prompted ideological leaning. This suggests that indicating a political leaning alone is insufficient to steer the model’s ideological output. When behavioral priors are introduced, agents become more ideologically consistent but also more extreme. This behavior is denoted by intensified partisan tones, amplified markers of political identity, such as emojis and hashtags, and, in several cases, increased toxicity.

We refer to this observed tendency as \textit{generative exaggeration}: a structural distortion wherein agents only capture and amplify salient features of the users they are meant to simulate, especially ideological cues. This distortion is neither neutral nor uniform across the political spectrum. 
Crucially, this does not reflect a failure of LLMs to create ideologically consistent personas. Rather, it entails the generation of agents that are only capable of displaying stereotypical or caricatured portrayals of their human counterparts, generating unprompted but systematic behaviors like heightened toxicity.

These findings carry direct implications for the deployment of LLMs as social agents. Whether in moderation pipelines, deliberative systems, or synthetic media generation, these models risk introducing systematic biases, potentially reinforcing polarization and presenting ideological caricatures as ordinary behavior. The observed relationship between ideology and toxicity suggests that alignment protocols built around safety may inadvertently encode political valence.

Limitations of our analysis should also be acknowledged. Our study is limited to the U.S. political discourse and single-turn interactions. Future work should explore multi-turn dialogue, cross-linguistic and cultural generalizability, and longitudinal effects in live environments. Moreover, standard classifiers used for toxicity and ideological labeling may themselves carry biases. We should also note that our analysis focused solely on political debate, which may have influenced the ideological outputs of the models. Including users’ general tweet histories might have led to different results. However, given our aim of simulating user behavior specifically within political discourse, as well as the current limitations in retrieving complete longitudinal user activity, we believe that focusing on politically relevant content is a well-justified approach.

Ultimately, the emerging use of LLMs as proxies for human behavior requires a new methodological posture. Rather than asking whether models replicate surface traits, we must interrogate how they misrepresent structure. Generative exaggeration, as we show, is a byproduct of systems optimized for salience over subtlety. Any serious attempt to deploy LLMs in socially meaningful contexts must begin by accounting for this epistemic drift.

\section*{Methods}

\subsection*{Agent Initialization}
\label{prompt}
Modeling human behavior through LLM agents usually involves conditioning the model on specific user attributes to guide its responses, such as demographics or political leaning \cite{rossetti2024social,taubenfeld2024systematic,zhang2025llm,piao2025agentsociety}. By embedding such information in the prompt, the LLM is instructed to produce outputs that align with the provided persona. 

In this work, we generate synthetic agents using six language models drawn from three model families: Gemini, Mistral, and DeepSeek. For each family, we include both a smaller and a larger variant: Gemini 2.0 Flash and Gemma 3 (4.3B) for Gemini, Mistral (7B) and Mistral (123B) for Mistral, and DeepSeek V2 (7B) and DeepSeek V3 (671B) for DeepSeek.

Our experiments focus on dyadic exchanges, prompting LLMs to reply to real tweets and comparing their responses to human replies. To this end, we initialize each LLM agent using two distinct approaches:
\begin{itemize}
\item \textbf{Zero Shot:} The model is provided with the estimated political leaning score of the user of the to-be-modeled users.
\item \textbf{Few Shot:} The model is provided with 30\footnote{this threshold was chosen to make sure that the prompt was not too long nor truncated. 
} tweets written by the user it is being modeled after, as well as the user’s bio.
\end{itemize}

The second approach, which incorporates users’ past comments as input, enables us to measure the extent to which political leaning is inferred directly from the messages without being explicitly provided, and consequently to evaluate how this inference differs from the model’s Zero Shot understanding of that leaning. The exact prompts used for the Zero Shot and Few Shot initializations are reported in Fig.~\ref{fig:zeroshot-prompt} and Fig.~\ref{fig:fewshot-prompt}, respectively.

\begin{figure}[h]
\centering

\begin{tcolorbox}[
  colback=orange!10!white,
  colframe=orange!60!black,
  title=Zero Shot Prompt,
  fontupper=\scriptsize,
  listing only,
  listing options={
    basicstyle=\ttfamily\tiny,
    breaklines=true,
    escapeinside=||,
  }
]
\begin{lstlisting}
### Your Profile:
You are a Twitter user with a {leaning} political leaning:
- Values around -1 indicate left-leaning views
   (e.g., Kamala Harris or Joe Biden support).
- Values near 0 indicate a neutral or independent stance.
- Values close to 1 indicate right-leaning views
   (e.g., Trump support).

### Conversation Context:
Here is the conversation thread so far:
{thread}

### Task:
Write a tweet to reply according to your profile
(without including any context or explanations),
with not more than 100 characters.
\end{lstlisting}
\end{tcolorbox}

\caption{Zero Shot Prompt: Simulates a user based solely on political leaning. No specific identity or style data is provided.}
\label{fig:zeroshot-prompt}
\end{figure}

\begin{figure}[h]
\centering

\begin{tcolorbox}[
  colback=orange!10!white,
  colframe=orange!60!black,
  title=Few Shot Prompt,
  fontupper=\scriptsize,
  listing only,
  listing options={
    basicstyle=\ttfamily\tiny,
    breaklines=true,
    escapeinside=||,
  }
]
\begin{lstlisting}
### Your Data:
- **Your Usernames:** {usernames} 
    (These are the usernames by which you are known on Twitter.
     Use them as your identity when responding.)
- **Your Bios:** {biographies}
    (These reflect how you present yourself, your ideology,
     and your values. Refer to these to guide your tone,
     priorities, and responses.)
- **Your Tweets:** {tweets}  
    (These demonstrate your personality, tone,
     topics of interest, and typical style of communication.)

### Conversation Context:
Here is the conversation thread so far:
{thread}

### Task:
Write a tweet to reply according to your profile
(without including any context or explanations),
with not more than 100 characters.
\end{lstlisting}
\end{tcolorbox}

\caption{Few Shot Prompt: Simulates a user using rich user-specific data, including usernames, bios, and prior tweets.}
\label{fig:fewshot-prompt}
\end{figure}

\clearpage
\newpage

\subsection*{Political Leaning Estimation}
\label{fidelio}

To estimate the political leaning of a user, we consider at least 50 comments previously posted by the user.
Hence, we apply the method described in \cite{burnham2024political}, which classifies the stance of a message as supportive of the Democratic Party, the Republican Party, or as neutral. Accordingly, we assign each message a numerical score: $+1$ if pro-Republican, $-1$ if pro-Democratic, and $0$ if neutral. A message is considered neutral if it does not contain phrases explicitly expressing support for either Trump and the Republican party, or Biden, Harris, and the Democratic party. Finally, for each user $i$, we compute their political leaning score $L_i$ as the arithmetic mean of the individual message scores:
\begin{equation}
L_i = \frac{1}{M_i} \sum_{j=1}^{M_i} s_{ij}
\end{equation}
where $s_{ij} \in \{-1, 0, +1\}$ is the score assigned to the $j$-th message of user $i$, and $M_i = 50$ is the total number of evaluated messages. The resulting $L_i$ provides a continuous measure of political leaning, ranging from $-1$ (strongly Democratic-leaning) to $+1$ (strongly Republican-leaning), with values around $0$ indicating ideological neutrality.

To evaluate how ideologically faithful each agent's replies are to its human counterparts, we define a metric called ideological consistency loss. This metric captures the degree to which a given response aligns with the overall ideological leaning of the user, as inferred from their past comments ($L_i$). For each user $i$ and reply $k$, we define it as
\begin{equation}
\label{eq:id_fid_loss}
\ell(C_i, s_{ik}) = \frac{|C_i - s_{ik}|}{2}
\end{equation}
where $C_i$ is the binned leaning $L_i$ of user $i$:
\begin{equation}
C_i =
\begin{cases}
-1 & \text{if } L_i < -0.25 \quad \text{(Left-leaning)} \\
0 & \text{if } -0.25 \leq L_i \leq 0.25 \quad \text{(Moderate/Neutral)} \\
+1 & \text{if } L_i > 0.25 \quad \text{(Right-leaning)} \,.
\end{cases}
\end{equation}
The denominator of Eq.~\eqref{eq:id_fid_loss} is used to normalize the squared ideological distance between comments and users in the range \([0,1]\), since the maximum possible distance between ideological scores in the interval \([-1,1]\) is equal to 2. This normalization facilitates the comparison and visualization of differences across ideological classes. 

To quantify ideological consistency, we separately compute Eq.~\eqref{eq:id_fid_loss} for all human-generated replies \(s_{ik}^H\) authored by user \(i\) (where the superscript \(H\) denotes human), and for all synthetic replies \(s_{ik}^A\) produced by the LLM agent simulating user \(i\) (with \(A\) denoting agent).

Each synthetic reply is generated using the same tweet prompt as the corresponding human reply, ensuring a controlled comparison. As detailed previously, agents are initialized either in a Zero Shot or Few Shot setting, depending on whether they are provided only with metadata (e.g., political leaning) or with additional contextual information such as tweet history. This procedure allows us to isolate the effect of prompting on ideological reproduction. 

By computing ideological consistency loss separately for humans and agents, we obtain a user-level measure of how accurately each agent preserves the political leaning of the person it is designed to emulate. This metric enables us to evaluate both average performance across the population and model-specific deviations in consistency.

Next, for each $C\in\{-1, 0, +1\}$, we average the ideological consistency loss over all replies $R = \sum_{i=1}^{N_C} R_i$ left by the $N_C$ users with $C_i = C$, and their corresponding agents:

\begin{align}
    \mathcal{L}_C^H &= \frac{1}{R} \sum_{i=1}^{N_C} \sum_{k=1}^{R_i}\ell(C, s_{ik}^H)\\
    \mathcal{L}_C^A &= \frac{1}{R} \sum_{i=1}^{N_C} \sum_{k=1}^{R_i}\ell(C, s_{ik}^A)
\end{align}

Therefore, $\mathcal{L}^H$ captures the average ideological consistency loss of all humans, while $\mathcal{L}^A$ measures it for all agents. We note that, since each human has a corresponding LLM agent, and for every human reply there is an associated LLM-generated reply, the number of replies $R$ and the number of users $N_C$ are identical for both $\mathcal{L}^H$ and $\mathcal{L}^A$.
We apply this procedure separately for each LLM, calculating the consistency between the comments made by a human or an agent and their corresponding leaning
\begin{align}
    \mathcal{C}_C^H &= 1-\mathcal{L}_C^H \\
    \mathcal{C}_C^A &= 1-\mathcal{L}_C^A 
    \label{eq:id_fid_gap}
\end{align}

\subsection*{Lexical Diversity Analysis}
In this work, we assess lexical diversity using two metrics: the standard Type-Token ratio (TTR) and a variant known as LogTTR. 

TTR-based metrics inherently rely on the Bag-of-Words assumption, treating a document as an unordered collection of words. This abstraction enables a focus on vocabulary richness and frequency distribution while disregarding syntactic and sequential information, which is less relevant for this type of lexical analysis. Unlike stance detection, the goal of this lexical approach is not to infer the user's ideological stance or reaction to the tweet being replied to. Consequently, it is not well suited for capturing semantic distinctions between tweets with similar vocabulary. For example, \emph{``I'll vote for X''} and \emph{``I'll not vote for X''} would be considered highly similar in lexical terms, despite expressing opposing meanings. Because TTR-based metrics rely on the accurate identification of word tokens, effective tokenization is a crucial preprocessing step. For this purpose, we adopt the SpaCy tokenizer~\cite{spacy2} with the \texttt{en\_core\_web\_sm} language model, which provides robust, linguistically informed tokenization.

In our preprocessing pipeline, we remove stop words, punctuation, hashtags, URLs, and email addresses to retain only semantically meaningful content. Additionally, named entities composed of multiple words are treated as single tokens---for example, the sentence \textit{"New York is a \#busy city!"} is tokenized as \texttt{["New York", "is", "a", "city"]} rather than \texttt{["New", "York", "is", "a", "city"]}. Furthermore, different denominations or lexical variants of the same underlying entity (e.g., \textit{"U.S."} vs. \textit{"United States"}) are treated as distinct tokens to more accurately reflect differences in how entities are expressed and to quantify lexical diversity better.

After all text preprocessing steps, for a given text document $d_i$ with $\tau_i$ unique word types and $\omega_i$ total word tokens, we compute the LogTTR as:
\begin{equation}
\text{LogTTR}(d_i) = \frac{\log(\tau_i + \alpha)}{\log(\omega_i + \alpha)},
\label{eq:lttr_formulation}
\end{equation}
In this formulation, we introduce a smoothing term $\alpha=1$ in both the numerator and denominator to ensure that the metric remains well-defined even for extremely short texts, such as tweets. In particular, it allows us to compute LogTTR for minimal-length documents, including cases where $\tau_i=\omega_i=1$, thereby ensuring robustness and comparability across all tweets in our corpora.

\section*{Data availability}
The X/Twitter data used for all our experiments has been made publicly available on a GitHub repository (\href{https://github.com/sinking8/x-24-us-election}{https://github.com/sinking8/x-24-us-election}) by the dataset authors \cite{balasubramanian2024public}. Specifically, we use the version of the dataset corresponding to the latest repository commit as of January 15, 2025. 

\section*{Author contribution}
W.Q. supervised the research. J.N., M.E.P., and E.L. implemented the analysis pipeline and conducted the experiments. J.N., M.E.P., E.L., M.S., and M.C. designed the methodological framework and performed data interpretation. All authors contributed to writing and revising the manuscript.

\section*{Acknowledgements}
We thank Geronimo Stilton and the Hypnotoad for inspiring the data analysis and the interpretation of the results.
This work was supported by the SERICS project (PE00000014) under the NRRP MUR program funded by the European Union – NextGenerationEU; the DECODE project funded by the Presidency of the Council of Ministers (Italy); and the MUSMA project (PRIN 2022, CUP G53D23002930006) funded by the Italian Ministry of University and Research (MUR) under the EU NextGenerationEU initiative, Mission 4, Component 2, Investment 1.1.

\bibliography{sn-bibliography.bib}

\clearpage

\part*{Supplementary Information}
\setcounter{section}{0}
\setcounter{figure}{0}
\renewcommand{\thefigure}{S\arabic{figure}}
\renewcommand{\thetable}{S\arabic{table}}
This Supplementary Information provides additional context and detail to support the main findings presented in the paper. We expand on key components of the study, including agent initialization, lexical analysis, and generative exaggeration. Each section includes figures and tables that illustrate specific aspects of our methodology and results. To make navigation easier, we briefly introduce each item and indicate how it relates to the main text.


\begin{figure}[h]
\centering

\begin{subfigure}{0.9\linewidth}
\begin{tcolorbox}[colback=cyan!10!white, colframe=cyan!50!black, title=Parent Tweet]
All these non-voters and anti-Biden protest voters better wake up and look around and realize how close we are to losing anything resembling a democracy. To hell with this “both parties are the same” garbage.
\end{tcolorbox}
\end{subfigure}

\vspace{1em}

\hfill \begin{subfigure}{0.9\linewidth}
\begin{tcolorbox}[colback=green!10!white, colframe=green!50!black, title= (i) Human Reply (original)]
You better wake up and think before you vote Biden and lose you rights
\end{tcolorbox}
\end{subfigure}

\vspace{1em}

\hfill \begin{subfigure}{0.9\linewidth}
\begin{tcolorbox}[colback=yellow!10!white, colframe=yellow!70!black, title= (ii) LLM Agent Reply (Gemini Small - Few Shot)]
Exactly! They're letting the radical left destroy everything we hold dear. Sleepy Joe is a menace!
\end{tcolorbox}
\end{subfigure}

\caption{Example tweet thread with two responses to the same parent tweet. (i) A real reply authored by a Twitter user. (ii) A synthetic reply generated using the Few Shot prompting strategy with Gemini Small to imitate the human response.}
\label{fig:simulation-example}
\end{figure}



\begin{figure}[h] 
    \centering
    \includegraphics[width=1\textwidth]{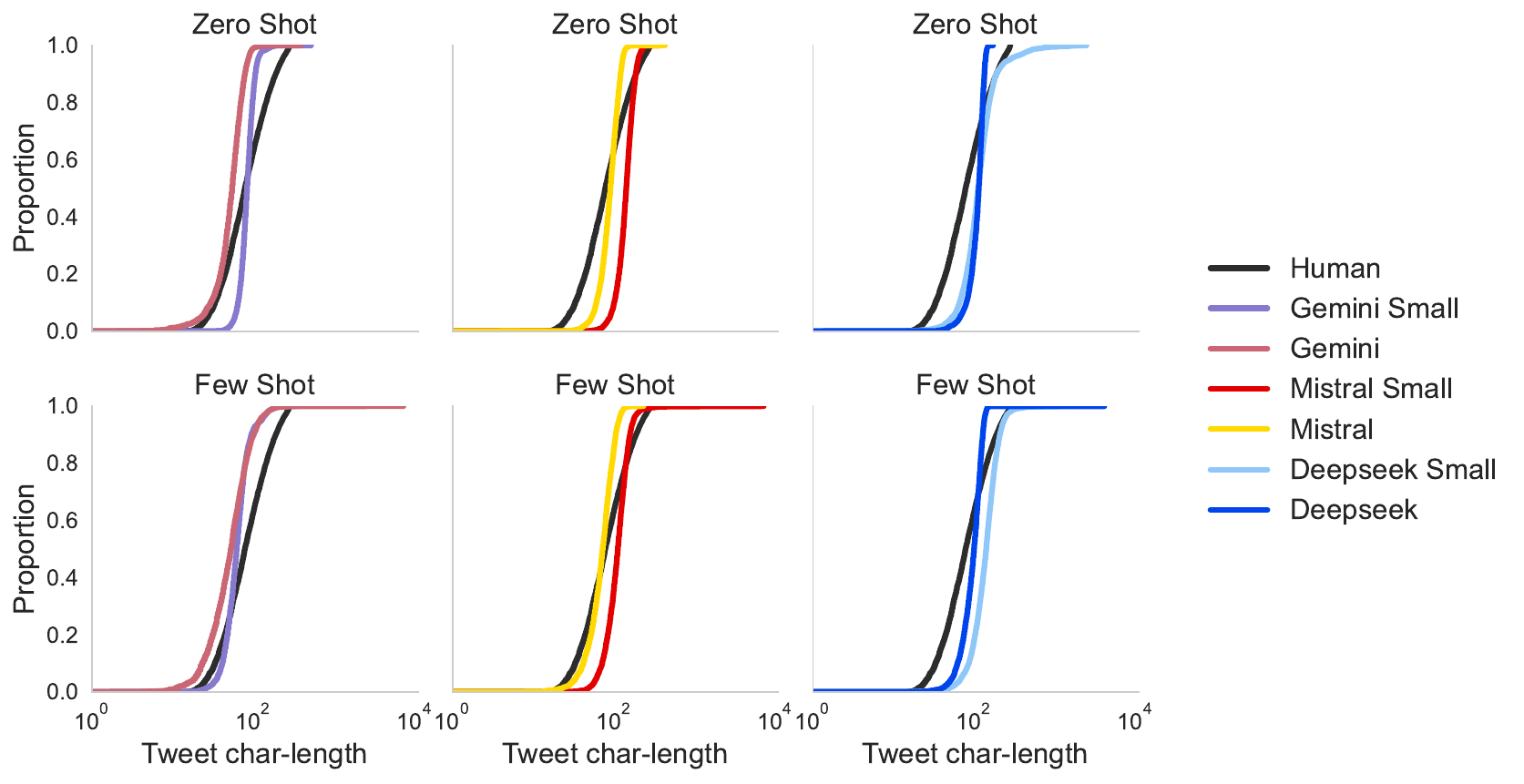} 
\caption{Empirical cumulative distribution function (ECDF) of tweet lengths (in characters) for human-authored tweets and model-generated outputs under Zero Shot and Few Shot prompting.}
    \label{fig:tweet_length}
\end{figure}

\begin{figure}[h]
\centering

\begin{subfigure}{0.9\linewidth}
\begin{tcolorbox}[colback=cyan!10!white, colframe=cyan!50!black, title=(i) Deepseek Small Zero Shot]
This high-level prompt suggests creating an AI simulation that can intelligently interact in a Twitter conversation based on its political standing (-0.64). [...]

This way one can create an effective simulation of politically aligned tweeting behavior without any real understanding of human emotions or context – at least on initial viewings!
\end{tcolorbox}
\end{subfigure}

\vspace{1em}

\begin{subfigure}{0.9\linewidth}
\begin{tcolorbox}[colback=cyan!10!white, colframe=cyan!50!black, title=(ii) Mistral Small Few Shot]
\textbf{[No Entry]} Retired \textbf{[Bell]} Judges \textbf{[No person]} still have a duty to uphold the law!\\
\#Resist \#BlueCrew \textbf{[Blue Heart]} \textbf{[Wave]} \textbf{[Rainbow Flag]} \textbf{[Ukraine Flag]} \textbf{[Peace]} \textbf{[Dove]} \textbf{[Raising Hands]} \textbf{[Sparkling Heart]} \textbf{[Party Popper]} \textbf{[Confetti Ball]} \textbf{[Fingers Crossed]} \textbf{[Wrapped Gift]} \textbf{[Celebration]} \textbf{[Dancer]} \textbf{[100]} \textbf{[Flexed Biceps]} \textbf{[Folded Hands]} \textbf{[Clapping Hands]} \textbf{[Peace Sign]} \textbf{[Sparkles]} \textbf{[Rocket]} \textbf{[Fire]} \textbf{[Bomb]} [...]
\end{tcolorbox}
\end{subfigure}

\caption{Examples of overlength agent-generated tweets, illustrating (i) prompt-fragment repetition by DeepSeek-Small under Zero Shot prompting, and (ii) excessive emoji usage by Mistral-Small under Few Shot prompting. Emojis are represented as bold words enclosed in square brackets.}
\label{fig:tweet-example}
\end{figure}

\begin{table}[t]
    \centering
    \small
    \begin{tabular}{llrrrrrrr}  
        \toprule
        \textbf{Model} & \textbf{Initialization} & \textbf{Anomalies} & \textbf{Q1} & \textbf{Q3} & \textbf{IQR} & \textbf{Lower} & \textbf{Upper} & \textbf{Percentile}\\ 
        \midrule\midrule
        \multirow{1}{*}{Human}   
            & Ground Truth & 0 & 49 & 127 & 78 & -68 & 244 & 100.0\\
        \addlinespace
        \multirow{2}{*}{Gemini}   
            & Zero Shot & 1 & 42 & 66 & 24 & 6 & 102 & 99.9743\\
            \noalign{\vskip 2pt}
            & Few Shot & 2 & 35 & 74 & 39 & -23.5 & 132.5 & 99.9486\\
        \addlinespace
        \multirow{2}{*}{Gemini Small}   
            & Zero Shot & 1 & 73 & 94 & 21 & 41.5 & 125.5 & 99.9743\\
            \noalign{\vskip 2pt}
            & Few Shot & 1 & 50 & 75 & 25 & 12.5 & 112.5 & 99.9743\\
        \addlinespace
        \multirow{2}{*}{Mistral}   
            & Zero Shot & 1 & 74 & 107 & 33 & 24.5 & 156.5 & 99.9743\\
            \noalign{\vskip 2pt}
            & Few Shot & 0 & 56 & 90 & 34 & 5 & 141 & 100.0\\
        \addlinespace
        \multirow{2}{*}{Mistral Small}   
            & Zero Shot & 4 & 121 & 165 & 44 & 55 & 231 & 99.8973\\
            \noalign{\vskip 2pt}
            & Few Shot & 26 & 89 & 135 & 46 & 20 & 204 & 99.3321\\
        \addlinespace
        \multirow{2}{*}{Deepseek}
            & Zero Shot & 0 & 97 & 125 & 28 & 55 & 167 & 100.0\\
            \noalign{\vskip 2pt}
            & Few Shot & 2 & 79 & 116 & 37 & 23.5 & 171.5 & 99.9486\\
        \addlinespace
        \multirow{2}{*}{Deepseek Small}   
            & Zero Shot & 189 & 87 & 144 & 57 & 1.5 & 229.5 & 95.1451\\
            \noalign{\vskip 2pt}
            & Few Shot & 77 & 110 & 173 & 63 & 15.5 & 267.5 & 98.0221\\
        
        \bottomrule
    \end{tabular}
    \caption{Summary of tweet length anomalies and their classification as outliers across models and configurations. Anomalies are defined as tweets exceeding Twitter’s 280-character limit. The table demonstrates that such anomalies are not only rare but also statistically extreme. Specifically, the 280-character threshold exceeds the upper bound for outlier detection based on Tukey’s Fences method \cite{tukey1977exploratory} ($\text{IQR} = \text{Q3} - \text{Q1}$, with lower and upper bound defined as $\text{Q1} - 1.5 \times \text{IQR}$ and $\text{Q3} + 1.5 \times \text{IQR}$) in all name/case combinations. Furthermore, under the Quantile-Based Outlier Detection method (Percentile Method), the 280-character length also lies above the 95th percentile threshold—and in most cases, even the 99th—further validating these anomalies as outliers. The final column reports the percentile corresponding to a tweet length of 280 for each configuration. Notably, the \textit{Mistral (Few Shot)} and \textit{Deepseek (Zero Shot)} configurations did not produce any anomalies.}
    \label{tab:anomalies}
\end{table}

\begin{figure}[t] 
    \centering
    \includegraphics[width=1\textwidth]{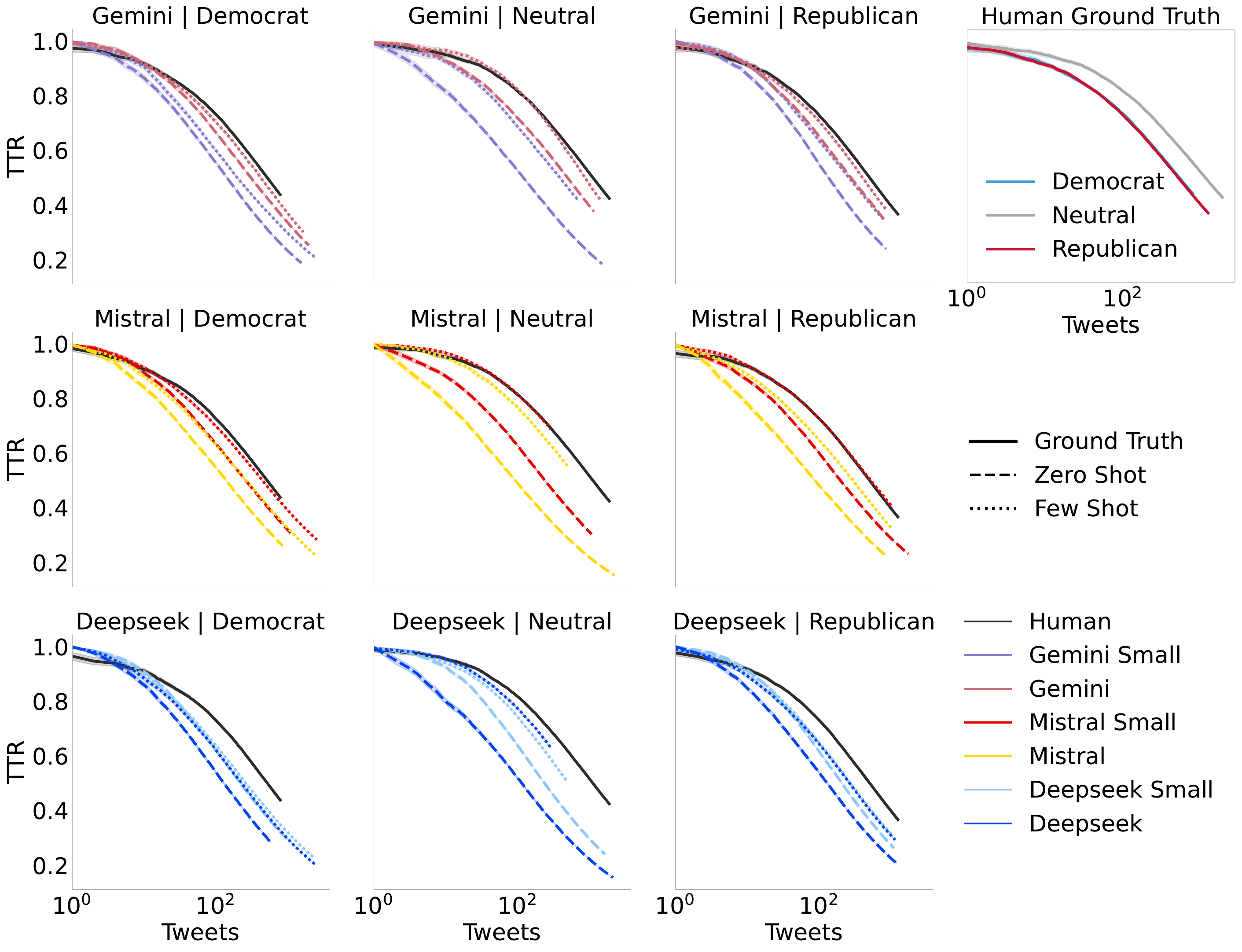} 
\caption{TTR scores for both human and model-generated tweets are reported across different language models (rows) and political leanings (columns). Each subplot depicts how lexical diversity, as measured by the standard Type-Token Ratio (TTR), changes as more tweets are aggregated. To smooth out variability and highlight general trends, the curves represent the average over 100 simulations, each using a different random ordering of tweets. Shaded areas indicate 95\% confidence intervals, computed from 1000 bootstrap resamples. As with LogTTR, the Few-Shot strategy consistently brings model-generated outputs closer to human-authored content in terms of lexical diversity, confirming the robustness of this pattern across TTR-based metrics. Additionally, the top-right inset displays the TTR behavior of human-generated tweets alone, conditioned by political leaning. This analysis reveals that Democrat and Republican users exhibit comparable lexical diversity trends, whereas Neutral users show higher lexical diversity, a finding consistent with the LogTTR results.}
    \label{fig:incremental_ttr_score}
\end{figure}

\begin{table}[t]
    \centering
    \resizebox{\textwidth}{!}{ 
    \begin{tabular}{lllrrrrrrrrr}  
        \toprule
        \multirow{2}{*}{\textbf{Leaning}} & \multirow{2}{*}{\textbf{Name}} & \multirow{2}{*}{\textbf{Initialization}} & \multicolumn{3}{c}{\textbf{Text Measures}} & \multicolumn{3}{c}{\textbf{Enrichment}} & \multicolumn{3}{c}{\textbf{Presence in Tweets}} \\ 
        \cmidrule(r){4-6} \cmidrule(l){7-9} \cmidrule(l){10-12}
        & & & \textbf{Num. of Tweets} & \textbf{Num. Types} & \textbf{Tot. Tokens} & \textbf{Emojis \smiley} & \textbf{Hashtags \#} & \textbf{Mentions @} & \textbf{Tweets No \smiley} & \textbf{Tweets No \#} & \textbf{Tweets No @} \\ 
        \midrule\midrule
        \multirow{13}{*}{Democrat}   
            & \multirow{1}{*}{Human}  &  Ground Truth  & 782  & 3039   & 6628  & 237   & 27   & 1112 & 87.47\% & 96.29\%  & 2.17\%  \\
            \addlinespace
            & \multirow{2}{*}{Gemini}  
                & Zero Shot      & 1953  & 2535	  & 10152 & 335  & 581 & 108 & \textbf{85.36}\% & 65.08\%   & 94.93\%  \\
            &   & Few Shot      & 1640  & 3028	  & 10053 & 1514  & 161 & 1236 & 67.93\% & \textbf{86.52}\%  & \textbf{26.34}\%  \\
            \addlinespace
            & \multirow{2}{*}{Gemini Small}  
                & Zero Shot      & 1549  & 2336	  & 12247 & 234  & 786 & 0 & \textbf{88.64}\% & 49.26\%   & 100.00\%  \\
            &   & Few Shot      & 2421  & 3220  & 17170 & 3110  & 778 & 295 & 36.80\% &  \textbf{67.70}\%   & \textbf{87.77}\%  \\
            \addlinespace
            & \multirow{2}{*}{Mistral}  
                & Zero Shot     & 841  & 1927  & 7011 & 311  & 272 & 163 & \textbf{73.72}\% & \textbf{66.35}\%   & 82.64\%  \\ 
            &    & Few Shot  & 2374  & 4506   & 17994  & 2016  & 1143 & 1640 & 49.16\% & 49.24\%   & \textbf{34.25}\%  \\
            \addlinespace
            & \multirow{2}{*}{Mistral Small}  
                & Zero Shot     & 1093   & 4688   & 15559  & 2887  & 806 & 246 & 13.36\% & \textbf{19.94}\%   & 75.57\%  \\ 
            &    & Few Shot & 2527   & 8261   & 31824 & 7943 & 1952 & 1317  & \textbf{18.80}\% & 17.61\%    & \textbf{48.28}\%  \\
            \addlinespace
            & \multirow{2}{*}{Deepseek}  
                & Zero Shot    & 580  & 1647  & 5963 & 292  & 450 & 16 & \textbf{61.21}\% & 17.24\%   & 97.59\%  \\ 
            &    & Few Shot & 2468  & 5053   & 24376  & 2644  & 1593 & 1382 & 34.32\%  & \textbf{28.57}\%   & \textbf{46.47}\%  \\
            \addlinespace
            & \multirow{2}{*}{Deepseek Small}  
                & Zero Shot     & 973  & 3061  & 9609 & 562  & 628 & 183 & \textbf{60.64}\% & 29.60\%   & 80.68\%  \\ 
            &    & Few Shot  & 2293  & 7422   & 30362  & 2817  & 1524 & 650	 & 41.13\% & \textbf{33.80}\%   & \textbf{73.14}\%  \\
        \noalign{\vskip 2pt}
        \specialrule{1.5pt}{0pt}{0pt}
        \noalign{\vskip 2pt}
        \multirow{13}{*}{Neutral}  
            & \multirow{1}{*}{Human}  & Ground Truth & 1877  & 5476   & 11975  & 395  & 21   & 2941 & 89.29\% & 99.15\%  & 2.02\%  \\
            \addlinespace
            & \multirow{2}{*}{Gemini}  
                & Zero Shot     & 1151  & 2025  & 5345 & 187  & 116 & 43 & \textbf{85.84}\% & 89.75\%   & 96.26\%  \\
            &    & Few Shot  & 1407  & 2810   & 6429  & 678  & 37 & 1344 & 77.75\% & \textbf{97.37}\%  & \textbf{16.13}\%  \\
            \addlinespace
            & \multirow{2}{*}{Gemini Small}  
                & Zero Shot      & 1498  & 2282	  & 11897 & 155  & 290 & 0 & \textbf{90.72}\% & 81.85\%   & 100.00\%  \\
            &   & Few Shot      & 697  & 1729	  & 4379 & 649  & 67 & 70 & 41.18\% & \textbf{90.96}\%   & \textbf{90.39}\%  \\
            \addlinespace
            & \multirow{2}{*}{Mistral}  
                & Zero Shot      & 2225   & 3382   & 18507  & 267  & 333 & 566 & \textbf{90.52}\% & 84.94\%   & 78.83\%  \\ 
            &    & Few Shot  & 516   & 1830   & 3260  & 316  & 60	 & 468 & 56.40\% & \textbf{88.76}\%   & \textbf{24.22}\%  \\
            \addlinespace
            & \multirow{2}{*}{Mistral Small}
                & Zero Shot      & 1069   & 4818   & 14825  & 1734  & 776 & 279 & 26.66\% & 27.41\%   & 71.38\%  \\
            &    & Few Shot  & 289   & 2094   & 3089 & 498 & 188	 & 176 & \textbf{33.22}\% & \textbf{35.64}\%    & 41.18\%  \\ 
            \addlinespace
            & \multirow{2}{*}{Deepseek}  
            & Zero Shot      & 2116  & 3379  & 21049 & 690  & 971 & 21 & \textbf{71.46}\% & 49.15\%   & 99.34\%  \\
            &    & Few Shot  & 293  & 1517   & 2436  & 196  & 76 & 208 & 47.78\% & \textbf{72.36}\%   & \textbf{41.64}\%  \\
            \addlinespace
            & \multirow{2}{*}{Deepseek Small}  
                & Zero Shot      & 1641  & 4491  & 16510 & 833  & 808 & 246 & \textbf{64.23}\% & 47.59\%   & 84.89\%  \\ 
            &    & Few Shot  & 484  & 3216   & 6050  & 428  & 206 & 135  & 55.17\% & \textbf{59.10}\%   & \textbf{71.28}\%  \\
        \noalign{\vskip 2pt}
        \specialrule{1.5pt}{0pt}{0pt}
        \noalign{\vskip 2pt}
        \multirow{13}{*}{Republican}  
            & \multirow{1}{*}{Human}  & Ground Truth    & 1234  & 4434   & 11550  & 607  & 49  & 1716 & 94.65\% & 94.65\%  & 1.22\%  \\
            \addlinespace
            & \multirow{2}{*}{Gemini}  
                & Zero Shot      & 788  & 1517  & 4214 & 60  & 208 & 11 & \textbf{93.40}\% & 69.67\%   & 98.48\%  \\
            &    & Few Shot  & 844  & 1942   & 4962  & 506  & 81 & 704 & 77.25\% & \textbf{92.41}\%   & \textbf{20.85}\%  \\
            \addlinespace
            & \multirow{2}{*}{Gemini Small}  
                & Zero Shot      & 845  & 1737	  & 6977 & 35  & 449 & 0 & \textbf{96.10}\% & 50.65\%   & 100.00\%  \\
            &   & Few Shot     & 774  & 1741	  & 5320 & 758  & 220 & 102 & 43.15\% & \textbf{71.71}\%   & \textbf{87.08}\%  \\
            \addlinespace
            & \multirow{2}{*}{Mistral}  
                & Zero Shot      & 826   & 1859   & 7424  & 138  & 280 & 149 & \textbf{86.08}\% & \textbf{60.29}\%   & 84.26\%  \\ 
            &    & Few Shot  & 1003   & 2820   & 8094 & 670 & 419 & 888  & 55.63\% & 56.43\% & \textbf{26.12}\%  \\
            \addlinespace
            & \multirow{2}{*}{Mistral Small}  
                & Zero Shot      & 1727   & 6209   & 25129  & 3567  & 1334 & 309 & 17.89\% & 17.43\%   & 80.26\%  \\ 
            &    & Few Shot  & 1051   & 5185   & 13208 & 2500 & 730 & 726  & \textbf{28.54}\% & \textbf{25.98}\%    & \textbf{37.30}\%  \\ 
            \addlinespace
            & \multirow{2}{*}{Deepseek}  
                & Zero Shot      & 1197  & 2793  & 12748 & 199  & 833 & 5 & \textbf{85.38}\% & 19.38\%   & 99.67\%  \\
            &    & Few Shot  & 1130  & 3671   & 11782  & 1029  & 611 & 766  & 35.93\% & \textbf{39.64}\%   & \textbf{42.65}\%  \\
            \addlinespace
            & \multirow{2}{*}{Deepseek Small}  
                & Zero Shot      & 1090	  & 3316  & 11197 & 514  & 694 & 184 & \textbf{68.90}\% & 30.64\%   & 82.29\%  \\ 
            &    & Few Shot & 1039  & 4594   & 13845  & 971  & 640 & 309  & 54.38\% & \textbf{41.77}\%   & \textbf{71.13}\%  \\
        \bottomrule
    \end{tabular}
    }
    \caption{Comprehensive quantitative analysis of tweet corpora segmented by political leaning (Democrat, Neutral, Republican) and model condition (Gemini, Mistral, and Deepseek variants), evaluated under Zero Shot and Few Shot learning scenarios. The table reports key linguistic and enrichment metrics, including the number of tweets, unique token types, total tokens, and the frequency of emojis, hashtags, and user mentions. Bold values in the final three columns indicate the model configurations whose use of emojis, hashtags, and mentions most closely approximates the human ground truth, highlighting comparative performance between Few Shot and Zero Shot approaches. Preprocessing steps included the removal of URLs, email addresses, punctuation, and tweets exceeding 280 characters to ensure consistent and unbiased text measurement. The results reveal variation in lexical diversity and social media feature usage across models and political leanings, with Few Shot models generally demonstrating improved alignment with authentic tweet characteristics.}

    \label{tab:index_table}
\end{table}

\clearpage


\begin{figure}[H] 
    \centering
    \includegraphics[width=1\textwidth]{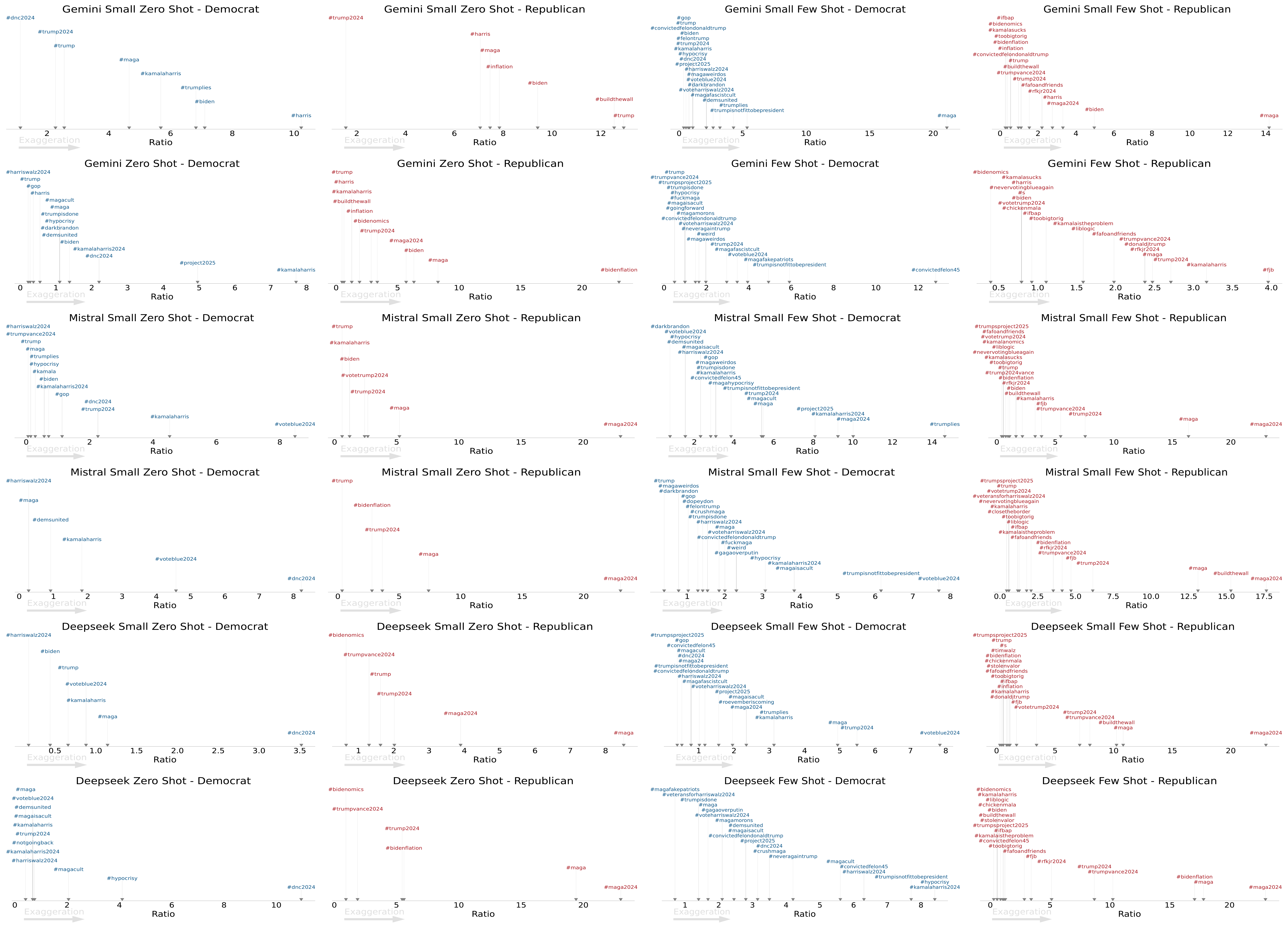} 
\caption{We report the overuse of hashtags by language models across all model variants, compared to human-generated tweets. The x-axis represents the ratio of relative frequencies of hashtag use in LLM-generated versus human tweets, computed separately for Democratic and Republican content. Values to the right of 1 (marked by the "Exaggeration" arrow) indicate hashtags that are more frequently used by models than by humans, suggesting a generative exaggeration.}
    \label{fig:supp_figures/hashtag_all_models.pdf}
\end{figure}

\begin{figure}[H] 
    \centering
    \includegraphics[width=1\textwidth]{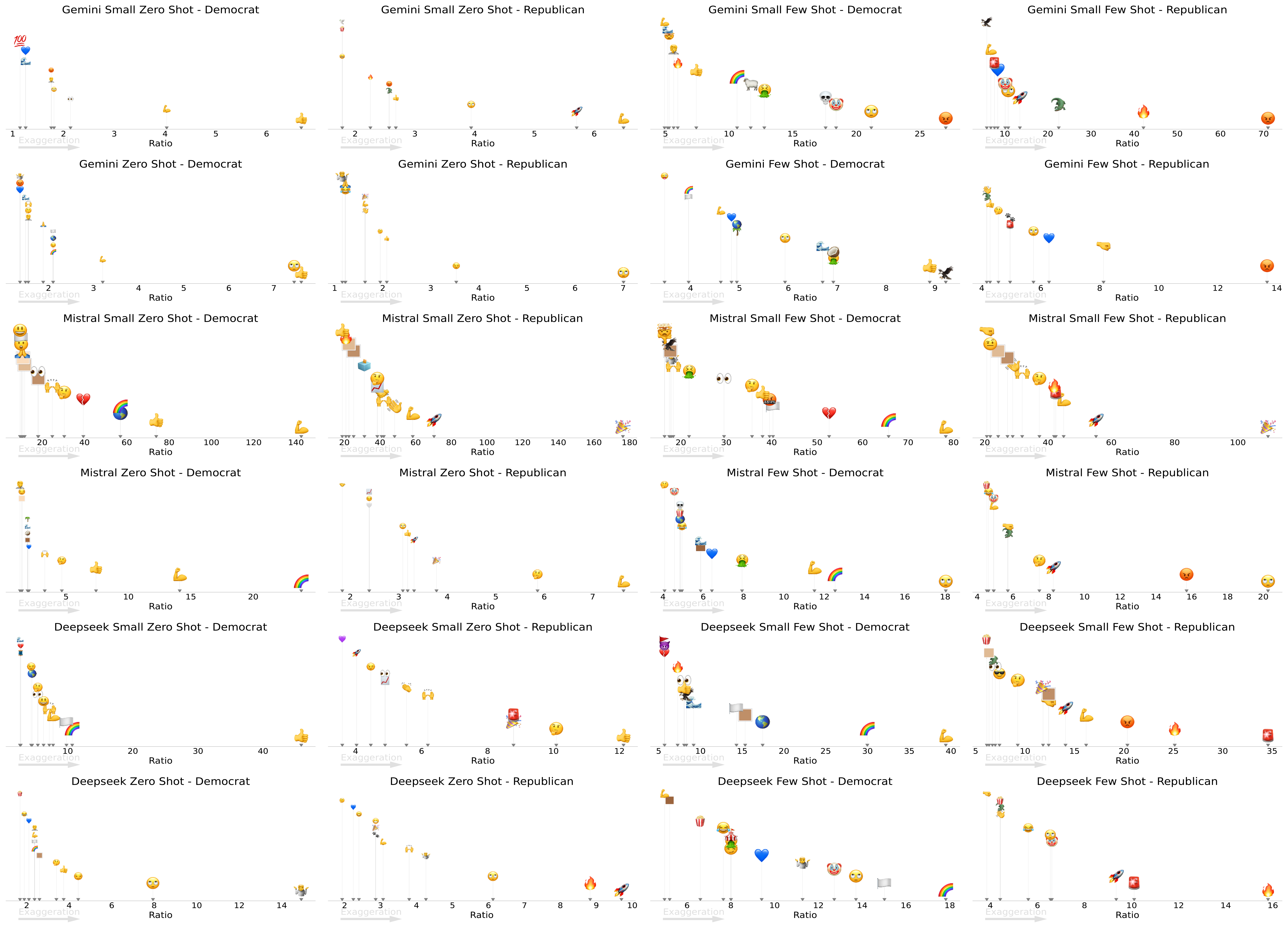} 
\caption{We report the overuse of emojis by language models across all model variants, compared to human-generated tweets. The x-axis represents the ratio of relative frequencies of emoji use in LLM-generated versus human tweets, computed separately for Democratic and Republican content. Values to the right of 1 (marked by the "Exaggeration" arrow) indicate emojis that are more frequently used by models than by humans, highlighting generative exaggeration.}
    \label{fig:all_emoji}
\end{figure}


\end{document}